\documentclass{article}\usepackage[]{graphicx}\usepackage[]{color}
\makeatletter
\def\maxwidth{ %
  \ifdim\Gin@nat@width>\linewidth
    \linewidth
  \else
    \Gin@nat@width
  \fi
}
\makeatother

\definecolor{fgcolor}{rgb}{0.345, 0.345, 0.345}

\usepackage{textcomp}

\usepackage{framed}
\makeatletter
 {\par\unskip\endMakeFramed%
 \at@end@of@kframe}
\makeatother

\definecolor{shadecolor}{rgb}{.97, .97, .97}
\definecolor{messagecolor}{rgb}{0, 0, 0}
\definecolor{warningcolor}{rgb}{1, 0, 1}
\definecolor{errorcolor}{rgb}{1, 0, 0}
\newenvironment{knitrout}{}{} 

\usepackage{alltt}

  \usepackage[margin=1in]{geometry}
  \usepackage{latexsym}
  \usepackage{graphicx}
  \usepackage{mathptmx}
  \usepackage{hyperref}
  \usepackage{mhchem}

\hypersetup{colorlinks=true,urlcolor=blue,citecolor=black,anchorcolor=black,linkcolor=black}

\makeatletter
\let\@internalcite\cite
\def\cite{\def\@citeseppen{-1000}%
    \def\@cite##1##2{(##1\if@tempswa , ##2\fi)}%
    \def\citeauthoryear##1##2##3{##1 ##3}\@internalcite}
\def\citeNP{\def\@citeseppen{-1000}%
    \def\@cite##1##2{##1\if@tempswa , ##2\fi}%
    \def\citeauthoryear##1##2##3{##1 ##3}\@internalcite}
\def\citeN{\def\@citeseppen{-1000}%
    \def\@cite##1##2{##1\if@tempswa, ##2)\else{}\fi}%
    \def\citeauthoryear##1##2##3{##1 (##3)}\@citedata}
\def\citeA{\def\@citeseppen{-1000}%
    \def\@cite##1##2{(##1\if@tempswa , ##2\fi)}%
    \def\citeauthoryear##1##2##3{##1}\@internalcite}
\def\citeANP{\def\@citeseppen{-1000}%
    \def\@cite##1##2{##1\if@tempswa , ##2\fi}%
    \def\citeauthoryear##1##2##3{##1}\@internalcite}
\def\shortcite{\def\@citeseppen{-1000}%
    \def\@cite##1##2{(##1\if@tempswa , ##2\fi)}%
    \def\citeauthoryear##1##2##3{##2 ##3}\@internalcite}
\def\shortciteNP{\def\@citeseppen{-1000}%
    \def\@cite##1##2{##1\if@tempswa , ##2\fi}%
    \def\citeauthoryear##1##2##3{##2 ##3}\@internalcite}
\def\shortciteN{\def\@citeseppen{-1000}%
    \def\@cite##1##2{##1\if@tempswa, ##2\else{}\fi}%
    \def\citeauthoryear##1##2##3{##2 (##3)}\@citedata}
\def\shortciteA{\def\@citeseppen{-1000}%
    \def\@cite##1##2{(##1\if@tempswa , ##2\fi)}%
    \def\citeauthoryear##1##2##3{##2}\@internalcite}
\def\shortciteANP{\def\@citeseppen{-1000}%
    \def\@cite##1##2{##1\if@tempswa , ##2\fi}%
    \def\citeauthoryear##1##2##3{##2}\@internalcite}
\def\citeyear{\def\@citeseppen{-1000}%
    \def\@cite##1##2{(##1\if@tempswa , ##2\fi)}%
    \def\citeauthoryear##1##2##3{##3}\@citedata}
\def\citeyearNP{\def\@citeseppen{-1000}%
    \def\@cite##1##2{##1\if@tempswa , ##2\fi}%
    \def\citeauthoryear##1##2##3{##3}\@citedata}
\def\@citedata{%
    \@ifnextchar [{\@tempswatrue\@citedatax}%
                  {\@tempswafalse\@citedatax[]}%
}

\def\@citedatax[#1]#2{%
\if@filesw\immediate\write\@auxout{\string\citation{#2}}\fi%
  \def\@citea{}\@cite{\@for\@citeb:=#2\do%
    {\@citea\def\@citea{, }\@ifundefined
       {b@\@citeb}{{\bf ?}%
       \@warning{Citation `\@citeb' on page \thepage \space undefined}}%
{\csname b@\@citeb\endcsname}}}{#1}}%

\def\@citex[#1]#2{%
\if@filesw\immediate\write\@auxout{\string\citation{#2}}\fi%
  \def\@citea{}\@cite{\@for\@citeb:=#2\do%
    {\@citea\def\@citea{, }\@ifundefined
       {b@\@citeb}{{\bf ?}%
       \@warning{Citation `\@citeb' on page \thepage \space undefined}}%
{\csname b@\@citeb\endcsname}}}{#1}}%

\def\@biblabel#1{}
\makeatother

\newdimen\bibindent
\newcommand{\Rstat}{\textsf{R}}
\newcommand{\ARMA}{\text{ARMA}}
\newcommand{\AR}{\text{AR}}
\IfFileExists{upquote.sty}{\usepackage{upquote}}{}
\begin{document}

\title{Betting and Belief: Prediction Markets and Attribution of Climate Change\thanks{Copyright \textcopyright\ 2016, IEEE.}}

\author{John J. Nay\thanks{Corresponding Author: john.j.nay@gmail.com and \href{http://johnjnay.com/}{johnjnay.com}.}, Martin Van der Linden,
Jonathan M. Gilligan
}

\maketitle
\section*{ABSTRACT}

Despite much scientific evidence, a large fraction of the American public doubts that greenhouse gases are causing global warming. We present a simulation model as a computational test-bed for climate prediction markets. Traders adapt their beliefs about future temperatures based on the profits of other traders in their social network. We simulate two alternative climate futures, in which global temperatures are primarily driven either by carbon dioxide or by solar irradiance. These represent, respectively, the scientific consensus and a hypothesis advanced by prominent skeptics. We conduct sensitivity analyses to determine how a variety of factors describing both the market and the physical climate may affect traders' beliefs about the cause of global climate change. Market participation causes most traders to converge quickly toward believing the ``true'' climate model, suggesting that a climate market could be useful for building public consensus.

\section{INTRODUCTION}

The climate change debate has become strongly polarized over the past two decades. Although the scientific consensus on the anthropogenic nature of climate change strongly increased, beliefs about climate change did not evolve much within the public \shortcite{Vandenbergh2013f}. In addition, the divide on anthropogenic climate change between liberals and conservatives has grown steadily as the question is becoming increasingly politicized and potentially disconnected from scientific evidence \shortcite{Kahan_2011}. The costs of misinformed climate policies are high. If climate change is not human-induced but is believed to be so, public resources will be spent on unnecessary efforts. On the other hand, if climate change is human-induced but not recognized to be so, the costs of inaction could be devastating. Effective climate policies require acting quickly, so it would be valuable to bring the public to a prompt and accurate consensus on the issue.

Attempts to foster such consensus face many social and psychological challenges, some of which could be addressed by creating climate prediction markets where participants can ``put their money where their mouths are'' \shortcite{Hsu2011,Vandenbergh2013f}. The idea of using prediction markets to efficiently aggregate information about uncertain event outcomes has been widely discussed \shortcite{Horn2014c}. Prediction markets have interesting theoretical properties \shortcite{Set1998,Hanson2012}, and perform well in terms of prediction accuracy and information aggregation in experiments \shortcite{Hanson2006,Healy2010} simulation models \shortcite{Klingert2012c,Jumadinova2011a} and the real-world \shortcite{Wolfers2006,pathak_comparison_2015,Dreber_2015}. However, to the best of our knowledge, the idea that prediction markets can generate consensus \emph{on the factors\/} affecting uncertain events has never been explored quantitatively.

\citeN{Bloch_2010} have proposed using derivatives markets to reduce the scientific uncertainty in estimating the impact of future climate change. Existing prediction markets (e.g. \href{https://hypermind.com/hypermind/app.html}{hypermind}, \href{https://www.betfair.com/exchange/}{betfair}, and \href{https://www.predictit.org/}{PredictIt}) focus on near term events such as elections months away, so it is difficult to extrapolate empirical findings to the climate case. Furthermore, we are interested in investigating the unobservable beliefs of traders. Therefore, we turn to simulation modeling informed by climate and economic theory. We simulate a prediction market where traders exchange securities related to climate outcomes to explore whether, and under what social and climate conditions, prediction markets may be useful for increasing convergence of climate beliefs. Our work can be extended as part of a computational design process for effective climate prediction markets and we release all our code along with this paper.

From a public policy perspective, changing the explanatory models of market participants is one of the most important roles that prediction markets might play. This is perhaps the most important social benefit of prediction markets given the predictive power of statistical supervised learning models, which could possibly provide information at a much lower cost than creating and maintaining a market \shortcite{goel_prediction_2010}. Effective climate policy not only requires an accurate consensus on future climate \emph{outcomes\/}. It also requires an accurate consensus on the causal \emph{mechanisms\/} influencing such outcomes. If people agree that temperature will rise, but some believe it will be due to greenhouse gases, while others believe that it will be caused by increased solar activity, inconsistent and ineffective policies may be implemented.

\section{RELATED WORK}

Agent-based simulations of prediction markets have been studied \shortcite{Klingert2012c,Tseng2010a,Jumadinova2011a}, including some that feature communication between agents. In these models, however, beliefs about the uncertain outcomes are constructed in rather abstract ways. In particular, beliefs are not based on structural models from which agents could derive causal implications, so these models are not suitable for investigating the convergence of the underlying explanatory models agents employ for prediction.

\shortciteN{Tseng2010a} created an agent-based model (ABM) of a continuous double auction market with multiple market strategies, including two variants of a zero-intelligence agent. They compared the behavior of their simulation with data from a prediction market for the outcome of political elections. They found that despite their simplicity, zero-intelligence agents capture some salient features of real market data. \shortciteN{Klingert2012c} compared the predictive accuracy of different kinds of simulated markets (continuous double auctions and logarithmic market scoring rules) and reached similar conclusions to experimental work by \shortciteN{Hanson2006}. None of these models featured agents learning and updating their beliefs.

\shortciteN{Jumadinova2011a} created a continuous double auction model in which agents update their beliefs about uncertain events based on newly-acquired information. The better the information set, the more likely the agents were to put higher weight on last period's prices when revising their beliefs. However, little structure was imposed on the information set and the way it was used to generate the weights for updating beliefs, so that model does not permit studying the convergence of trader beliefs about predictive models. \shortciteN{Ontanon2009} studied the effect of deliberation on a simulated prediction market, in which agents used case-based reasoning \shortcite{Aamodt1994} to debate uncertain outcomes with their neighbors in a social network. This model could in principle be used to assess convergence of predictive models, but this was not done and the process of forming and revising beliefs through argumentation is not well suited to studying stochastic time series, such as those relevant to climate prediction markets.

\section{MODEL DESIGN}

In our model, traders bet on global temperature anomalies six years in the future.
During the six-year period, traders buy and sell futures.
Every year, traders update their models and forecast of future temperatures based on newly available data.
At the end of each six-year period, winners collect gains and traders revise beliefs about climate models, based on their ideology and the beliefs of top earners in their social network.
In this section, we describe the models used to generate future temperature data, agents beliefs about those models, the market procedures, and the social network connecting agents.
We conclude with details on model dynamics and an overview of the model parameters we experimentally vary. A full Overview, Design concepts and Details (ODD) specification can be found on \href{https://github.com/JohnNay/predMarket}{the project website}.

\subsection{Temperature Models}
\label{sec:temp.models}

For climate time-series, we use the annual anomaly of global mean temperature. For years from 1880--2014 we use the GISTEMP global mean land-sea annual temperature anomalies \shortcite{GISTEMP_2016,Hansen_2010} and for years from 2015 onward, we project future climates under two alternative theories: In both theories, changes in global temperature are proportional to changes in a deterministic forcing plus a stochastic noise term. For simplicity, we choose two alternative expressions for the deterministic climate forcing: one, which corresponds to conventional climate science, takes the natural logarithm of the atmospheric carbon dioxide concentration in parts per million \shortcite[p.~37]{Archer_2012}, and the other, which corresponds to an alternative theory advocated by many who doubt or reject conventional climate science, takes the total solar irradiance (the brightness of sunlight, in Watts per square meter, at the top of the atmosphere) averaged over the 11-year sunspot cycle \shortcite{Soon_2005}.  Most scientific models of the earth's climate include many forcings, including carbon dioxide, other greenhouse gases, aerosols, total solar irradiance, and more. In these models, the changing $\ce{CO2}$ concentration is, by a large margin, the strongest single forcing \shortcite[p.~14]{IPCC_WG1_AR5_2013}. Choosing only one forcing term for each of the competing models simplifies comparison because each model has the same number of adjustable parameters, therefore we consider only \ce{CO2}.

For $\ce{CO2}$ we used historical emissions through 2005, harmonized with RCP 8.5 representative concentration pathway from 2005 onward \shortcite{rcp_database_2009,Riahi_2011} and for TSI we used the harmonized historical values, with a projection through 2100 from \shortciteN{Velasco_Herrera_2015}, which is, to our knowledge, the only prediction of TSI for the entire 21\textsuperscript{st} century.

Warming coefficients for each model were determined by linear regression of historical temperatures from 1880--2014 against the historical values for each model's forcing term. The noise model was determined by fitting an $\ARMA(p,q)$ model to the residuals from the regression, using the \href{http://mc-stan.org}{Stan} probabilistic modeling language and the \href{https://cran.r-project.org/web/packages/rstan/}{\texttt{rstan}} package \shortcite{Stan2016}. Stan proved more numerically stable than the \Rstat\ \href{http://CRAN.R-project.org/package=nlme}{\texttt{nlme}} package
for fitting ARMA noise models. We identified the optimal model for the auto-correlated noise term by performing the regression analysis for all combinations of $p, q \in \lbrace 0,1,2 \rbrace$ and using the Widely Applicable Information Criterion to select the optimal noise model \shortcite{Watanabe_2013,Gelman_2014}. In both cases (TSI and $\log\ce{CO2}$), the optimal noise model was $\AR(1)$.

Future climates were generated by applying the future climate forcings to Eq.~\ref{eq:climate}:
\begin{equation}
T_{\text{model}}(t) = \beta_{\text{model}} F_{\text{model}}(t) + \varepsilon,
\label{eq:climate}
\end{equation}
where $T_{\text{model}}(t)$ is the temperature at time $t$, under a given model of what causes warming,
$F_{\text{model}}(t)$ is the forcing (TSI or $\log\ce{CO2}$) at time $t$,
$\beta_{\text{model}}$ is the regression coefficient, and
$\varepsilon$ is a noise term.
The coefficient $\beta$ and the parameters of the ARMA noise model were fit to the historical data (1880--2014).
An example realization of future climates for both the $\ln(\ce{CO2})$ and TSI models is shown in Fig.~\ref{fig:climateplot}.

\begin{knitrout}
\definecolor{shadecolor}{rgb}{0.969, 0.969, 0.969}\color{fgcolor}\begin{figure}

{\centering \includegraphics[width=\maxwidth]{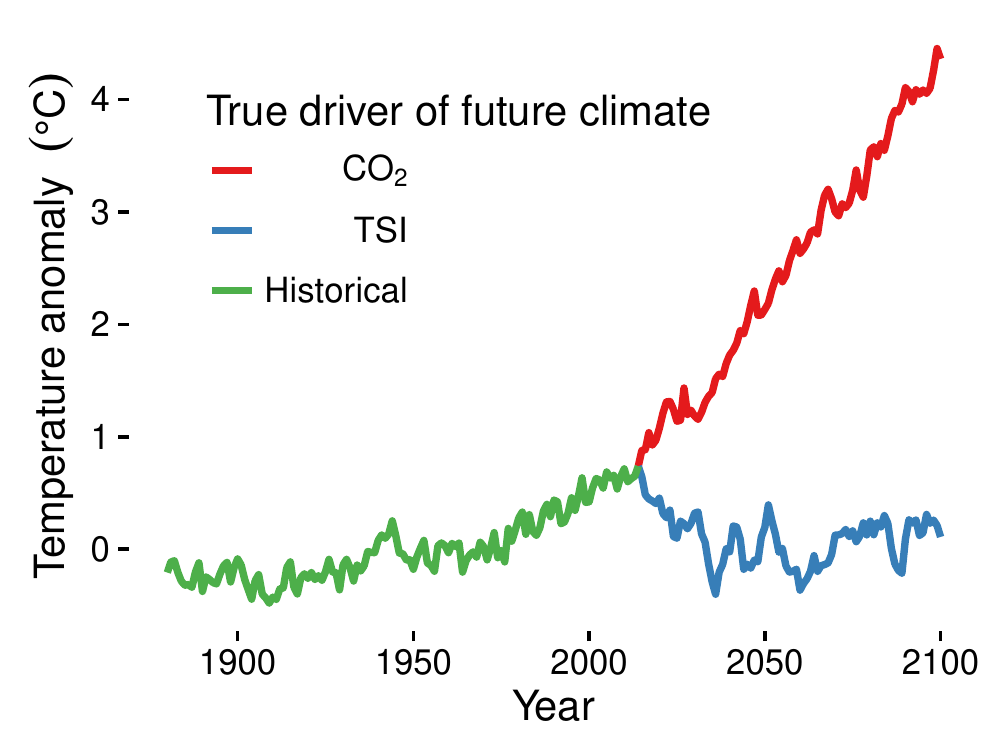} 

}

\caption[Historical measurements of temperature and a realization of possible future temperatures under two alternate models of climate physics]{Historical measurements of temperature and a realization of possible future temperatures under two alternate models of climate physics.}\label{fig:climateplot}
\end{figure}

\end{knitrout}

\subsection{Climate Beliefs}

Traders use one of the two models (temperature depends on \ce{CO2} or TSI) to forecast future temperature.
These models are interpreted as the trader's beliefs about the true climate process, the driving factor of long-term global temperatures.
They represent pervasive positions on climate change in the public debate. In order to approximately match the current configuration of beliefs in climate change in the United States, during model initialization, the \ce{CO2} model is randomly assigned to half of the traders, while the TSI model is assigned to the other half of the traders. These random model assignments are made \emph{before} the market initially opens.

Both when the true data generating process is \ce{CO2} and TSI, at model initialization approximately half of the traders use the true data-generating model to make predictions.
Traders using the true model do not necessarily make perfectly accurate predictions however.
Although these traders believe in the correct \emph{functional form\/} of the model, they still need to calibrate their model based on limited noisy data.
Therefore, the values these traders assign to the parameters of the model will typically be different from the exact parameters in the data-generating process.

\subsection{Traders and Markets}

Traders are initially endowed with a single experimental currency unit (ECU).
Traders use their model to forecast the distribution of future temperature and determine their reservation price for different securities. Each security pays 1 ECU at the end of the trading sequence if the temperature at the end of the sequence falls into a certain range.

Traders are risk-neutral expected utility maximizers.
Therefore, their reservation price for a security is simply their assessment of the probability that the temperature will fall in the range covered by the security at the end of the sequence.
At each time-step (year), the agents use the new year's temperature data to re-estimate the coefficients for the model they believe explains climate change, using Bayesian linear regression with an $\AR(1)$ noise model. Traders use the joint posterior probability distribution of regression and noise coefficients to estimate the probability distribution for the  temperature at the end of the current trading sequence. Stan was sufficiently fast that we could perform a full Bayesian analysis at each time step.

Traders then use this posterior probability distribution to assign reservation prices for buying and selling securities.
Based on their reservation price, agents behave as ``zero-intelligence'' traders \shortcite{Gode1993}.
They attempt to sell securities at a random price above their reservation, and to buy securities at a random price below their reservation.
These trading strategies are simple but provide accurate approximations of behavior in prediction markets \shortcite{Klingert2012c}, and in financial markets more broadly.

Based on traders' sell and buy orders, traders exchange securities following a continuous-double auction (CDA) procedure (see Model Dynamics below for more details).
CDAs or some close variants are common procedures to match buy and sell orders.
CDA are notably used on large stock markets \shortcite{Tseng2010a}.

\subsection{Social Network}

Traders are part of a social network where each agent forms two links at random, and then forms links randomly, ensuring that each agent is connected to at least two other agents.
Every time securities are realized, each trader looks at the performance of her richest neighbor in the network.
Traders start with the same initial amount of ECU and differences in ECU can only come from market interactions.
Therefore, if some trader is poorer than her richest neighbor, the trader interprets it as a signal that her richest neighbor may have a better model of the climate.
Then, the trader considers adopting the model of her richest neighbor.
For each trader, the willingness to revise her belief is determined by how ideologically loaded her belief is \shortcite{Kahan_2011}, which is a parameter we vary in our experiments.

\begin{figure}[t]
\begin{center}
\includegraphics[width=0.50\textwidth]{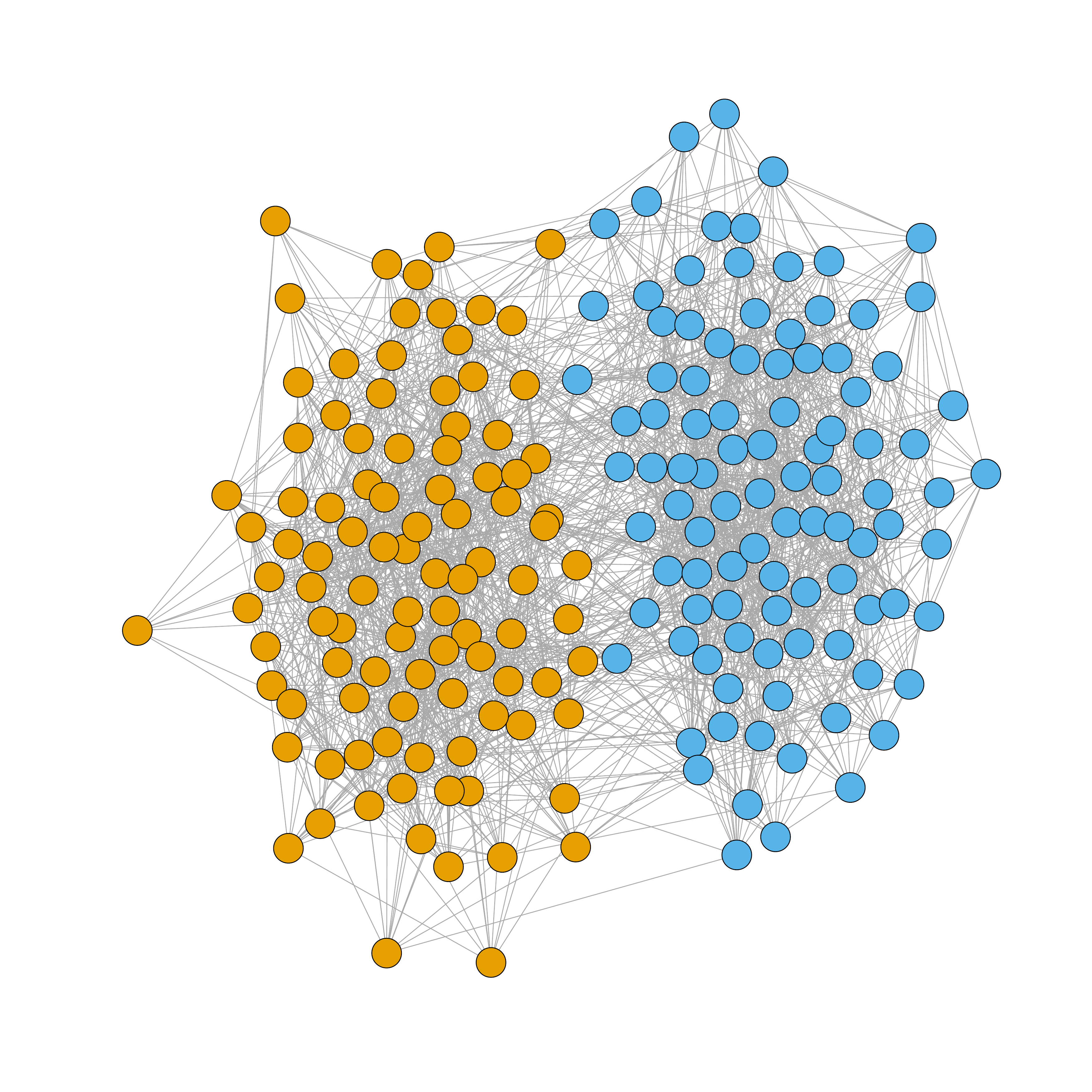}
\caption{An example of the social network among traders. Yellow and blue circles represent traders who believe in the two different models of climate change. Lines indicate social network connections. In a highly segmented network (\texttt{seg} close to 1), most links connect like-minded traders and few connect traders with opposing beliefs.}\label{fig:social_network}%
\end{center}
\end{figure}

An example of a snapshot of a social network is depicted in Fig.~\ref{fig:social_network}.
A segmentation parameter controls the homophily of the network: the extent to which traders are preferentially linked with other traders who share their initial belief in the cause of climate change (\ce{CO2} or TSI), as opposed to traders with the opposite intial belief. This parameter varies from 0 (no preference for like-minded traders) to 1 (traders are only connected to like-minded traders).
Although traders can change their beliefs over time, the connections between traders do not change as the market unfolds, i.e the edges are fixed.
We vary the segmentation parameter in our experiments.

\subsection{Model Dynamics}

The time periods $t$ are grouped into trading \emph{sequences\/}.
In a given sequence, the potential payments associated with traded securities are all based on the temperature at the end of the sequence.
For instance, the third trading sequence might start in period $t = 1964$ and end in period $t = 1970$.
In this case, a security traded in the third sequence pays 1 ECU if the temperature at $t = 1970$ falls into the range of temperatures covered by the security.

At each time $t$, traders are assumed to know the past value of the temperature ${T}_{0:t}$, carbon dioxide $\text{CO2}_{0:t}$, and total solar irradiance $\text{TSI}_{0:t}$.
In a sequence finishing at time $t^*$, traders also have common knowledge of $\text{CO2}_{t:t^*}$ and $\text{TSI}_{t:t^*}$, the future values of carbon dioxide and total solar irradiance up to $t^*$.
However, at any $t$, traders do not know the value of any future temperatures.
Thus, the traders know the forcing terms for the rest of the trading sequence, but do not know the value of ${T}_{t^*}$.
Traders can only predict ${T}_{t^*}$ using their approximate model and their knowledge of ${T}_{0:t}$, $\text{CO2}_{0:t^*}$, and $\text{TSI}_{0:t^*}$.
Notice that because ${T}_{0:t}$, $\text{CO2}_{0:t^*}$, and $\text{TSI}_{0:t^*}$ are common knowledge, in each period $t$, any two traders with the same approximate model $m$ form the same stochastic beliefs about future temperatures $P_{t,m}({T}_{t^*} ~|~ {T}_{0:t}, \text{CO2}_{0:t^*}, \text{TSI}_{0:t^*})$.
The probability distribution $P_{t,m}$ incorporates both epistemic uncertainty (the trader does not know the true values of the coefficients of the climate model) and aleatory uncertainty (in addition to deterministic warming, the temperature exhibits stochastic noise that is directly modeled).

 At each time $t$, traders:
\begin{enumerate}
\item  re-calibrate $m$, their approximate model at time $t$, based on the new set of temperature data available at $t$,
\label{step:calibrate}
\item  use the posterior probability distribution from (\ref{step:calibrate}) to
assign beliefs about the probability distribution of future temperatures at time $t^*$:
$P_{t,m}({T}_{t^*} ~|~ {T}_{0:t}, \text{CO2}_{0:t^*}, \text{TSI}_{0:t^*})$
and use it to determine the expected value they attach to each security, and
\item  trade on the CDA market as follows:
  \begin{itemize}
    \item Every trader $i$ chooses at random a security $s_i^B$ she will try to buy.
    \item Every trader $i$ also chooses at random a security $s_i^S$ she will try to sell among the securities she owns a positive amount of (if any).
    \item Traders then decide of their selling price $p_i^S$ and buying price $p_i^B$.
  To do so, traders first compute their expected values $E(s_i^B)$ and $E(s_i^S)$ for securities $s_i^B$ and $s_i^S$ (where expected values are with respect to $i$'s approximate model at time $t$).
  Then traders set $p_i^S$ at random above $E(s_i^S)$ and $p_i^B$ at random below $E(s_i^B)$ (see Model Parameters below for more details).
    \item Traders go to the market one at the time, in an order drawn randomly for each $t$.
    \item When trader $i$ comes to the market, she places limit orders in the order book.
   These orders specify that $i$ is willing to buy $s_i^B$ at any price below $p_i^B$, and to sell $s_i^S$ at any price above $p_i^S$.
    \item The market maker attempts to match $i$'s orders with some order which was put in the book before $i$ came to the market.
    \item If there are outstanding sell offers for $s_i^B$ at a price below $p_i^B$, a trade is concluded.
   Trader $i$ buys one unit of $s_i^B$ from the seller who sells at the lowest price below $p_i^B$, and the sell and buy offers are removed from the order book.
    \item If there are outstanding buy offers for $s_i^S$ at price above $p_i^S$, a trade is concluded.
   Trader $i$ sells one unit to the buyer who buys at the highest price above $p_i^S$, and the sell and buy offers are removed from the order book.
    \item When all traders have come to the market, any remaining outstanding offer is removed from the order book, and the trading period is concluded.
  \end{itemize}
\end{enumerate}

At $t^*$, when the sequence ends, there is only one security $s^*$ associated with a range of temperatures including the actual temperature $T_{t^*}$.
At $t^*$, traders:

\begin{enumerate}
  \item  receive 1 ECU per unit of $s^*$ they own, and
  \item consider adopting their neighbors' approximate model as described in the behavioral parameters sub-section below.
\end{enumerate}

\subsection{Model Parameters}

The model depends on the following parameters, which we vary in simulation experiments to determine their effects on the convergence of beliefs. We group the parameters into climate, network and individual behavioral factors.
\begin{itemize}
  \item Climate parameters:
  \begin{itemize}
    \item  \texttt{true.model}: temperature data-generating process (\ce{CO2} or TSI).
  \end{itemize}

  \item Network parameters:
  \begin{itemize}
    \item  \texttt{n.traders}: the number of traders.
    \item  \texttt{n.edg}: the number of edges in the social network.
    \item  \texttt{seg}: the segmentation parameter for the social network.
  \end{itemize}

  \item Behavioral parameters:
  \begin{itemize}
    \item  \texttt{risk.tak} : determines the distribution of risk tolerance with respect to successfully buying or selling securities.     Higher risk tolerance corresponds to demanding more aggressive prices for buying and selling, and hence, a higher risk of not consummating a trade that would be mutually advantageous to the buyer and seller.

    For each trader $i$, the level of $\texttt{risk.tak}_i$ is drawn uniformly at random from $[0,\texttt{risk.tak}]$.
    The higher $\texttt{risk.tak}_i$, the higher the price $i$ will demand for selling securities, and the lower the price $i$ will offer for buying.

    Formally, in each period, trader $i$ picks her buying or selling price for $s$ uniformly at random in the interval
    $[(1 - \texttt{risk.tak}_i) \texttt{reserv}_{i}, \allowbreak  \texttt{reserv}_{i,t}]$ for buying and
    $[\texttt{reserv}_{i}, \allowbreak (1  + \texttt{risk.tak}_i) \texttt{reserv}_{i,t}]$ for selling.

  \item  \texttt{ideo} :  determines the degree of ``ideology'' of traders.
    For each trader $i$, the level of $\texttt{ideo}_i$ is drawn uniformly at random from $[0,\texttt{ideo}]$.
    If \texttt{ideo} is high, traders will not revise their approximate models easily, even when faced with evidence that their richest neighbor is doing better than them.
    Formally, for each trader $i$ and each sequence, $\texttt{ideo}_i$ is the probability that $i$ adopts the approximate model of her richest neighbor if that neighbor is doing better than $i$ at the end of the sequence (in monetary terms).
  \end{itemize}
\end{itemize}

\section{RESULTS}

\subsection{Historical Climate}

As a validation that the market is operating correctly, we ran the market using actual historical temperatures from 1880 to 2014, with market betting from 1931 to 2014. Greenhouse gas concentrations did not become high enough to begin dominating natural climate variations until the 1970s or so, and it took until the 1990s for the temperature record to show clear signs of anthropogenic interference \shortcite{IPCC_WG1_AR5_2013}. Consistent with this, our simulated historical trading sequence (Fig.~\ref{fig:time}) shows convergence to belief in the TSI model until the early 1970s, after which traders converge toward believing the \ce{CO2} model.
\begin{knitrout}
\definecolor{shadecolor}{rgb}{0.969, 0.969, 0.969}\color{fgcolor}\begin{figure}[t]

{\centering \includegraphics[width=\maxwidth]{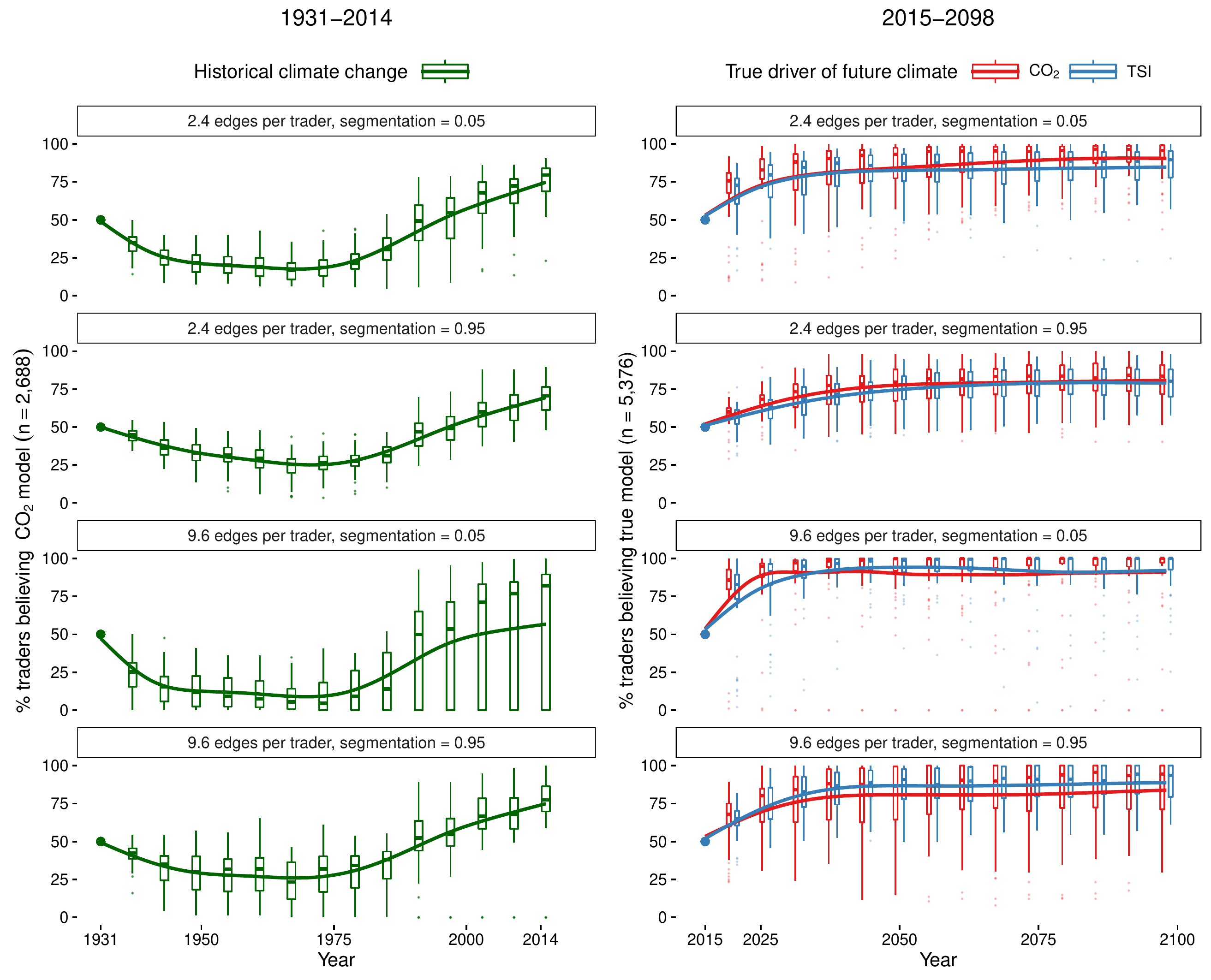} 

}

\caption[Convergence over trading sequences for different degrees of social-network connection and segmentation]{Convergence over trading sequences for different degrees of social-network connection and segmentation.}\label{fig:time}
\end{figure}

\end{knitrout}

\subsection{Future Scenarios}

Our primary focus is on simulating the market in future scenarios. This is more relevant for policy design and more interesting theoretically due to the increasing divergence of global temperature values under the two future scenarios. The sensitivity analysis is based on a Latin hypercube sampling of 500 parameter sets from the following distributions \shortcite{beachkofski_improved_2002,carnell_lhs_2012}.

\begin{itemize}
  \item \texttt{ideo} $\sim \text{Uniform}(0,1)$
  \item \texttt{n.edge} $\sim \text{Uniform}(100,200)$ (mapped into integer)
  \item \texttt{n.traders} $\sim \text{Uniform}(50,250)$ (mapped into integer)
  \item \texttt{risk.tak}  $\sim \text{Uniform}(0,1)$
  \item \texttt{seg} $\sim \text{Uniform}(0,1)$
  \item \texttt{true.model} $\sim \text{Bernoulli}(0.5)$
\end{itemize}

The future \ce{CO2} forcing is taken from the RCP~8.5 scenario \shortcite{Riahi_2011} and the future TSI forcing is taken from a prediction of 21\textsuperscript{st} Total Solar Irradiance by \shortciteN{Velasco_Herrera_2015}. Because of the stochastic noise term, the temperature time series were different in each simulation.

We used the model to perform 10 full simulations for each of the 500 input parameter sets and average the 10 convergence scores. We conducted multiple simulations for the same parameter set because there is stochasticity in the tempertature time series, the social network structure, and the agent decision models. We then conducted a partial rank correlation coefficient analysis on the relationship between the input matrix, $X$, and the resulting simulated outcome vector of mean belief convergence scores, $y$ \shortcite{Marino2008,pujol_sensitivity:_2014,saltelli_sensitivity_2009}. Partial correlation computes the linear relationship between the part of the variation of $X_i$ and $y$ that are linearly independent of other $X_j$  ($j \neq i$).  The difference between the partial correlation and the partial rank correlation that we use here is that we first rank-transform that data in order to capture potentially non-linear relationships. We conduct 1,000 bootstrapped estimations of the partial rank correlation coefficients to obtain 95\% confidence intervals.

\begin{knitrout}
\definecolor{shadecolor}{rgb}{0.969, 0.969, 0.969}\color{fgcolor}\begin{figure}[t]

{\centering \includegraphics[width=\maxwidth]{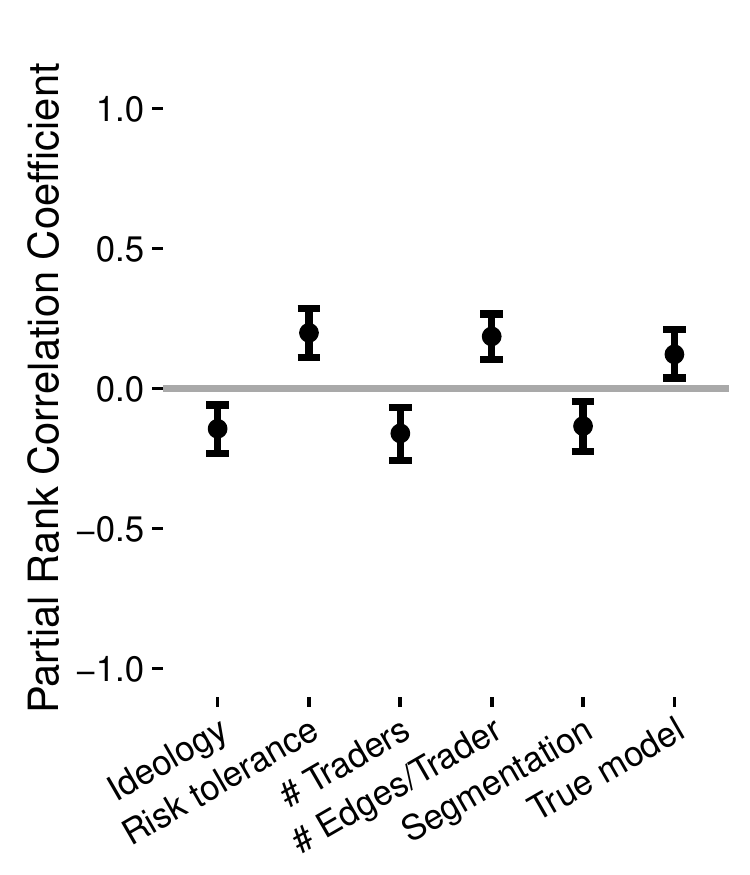} 

}

\caption[Estimated effects of model parameters on convergence of beliefs in future scenario]{Estimated effects of model parameters on convergence of beliefs in future scenario. Positive numbers accelerate convergence and negative numbers retard it. True model refers to the true climate model being \ce{CO2}, so traders converge to the true model more quickly when carbon dioxide drives climate change than when total solar irradiance does. Bars represent 95\% confidence intervals.}\label{fig:sa}
\end{figure}

\end{knitrout}

The sensitivity analysis averages over time and thus masks time trends. We randomly drew from the above distributions for ideology, risk tolerance, and the number of traders, and conducted an experiment crossing 95\textsuperscript{th} and 5\textsuperscript{th} percentile values for the number of edges per trader and the segmentation of the social network and both values for the true model. We collected the time series of belief-convergence across these eight designs to visualize the distributions of convergence over time (Fig.~\ref{fig:time}). Under most parameterizations, the median fraction of traders believing in the true model reaches 75\% in 10--20 years.

\section{DISCUSSION}

Our sensitivity analysis, Fig.~\ref{fig:sa}, shows that ideology, risk tolerance, the number of traders, the number of edges per trader, the segmentation of the social network, and the true model all statistically significantly affect convergence.  Increasing the number of edges per trader increases the flow of information through the market, which causes traders to converge toward believing the ``true'' climate model.

Segmented social networks reduce convergence by creating an ``echo-chamber'' effect, in which lack of interaction between traders with different views reduces traders' access to information that could persuade them to change their beliefs. This is apparent in Fig.~\ref{fig:time}, where the rate of convergence in highly segmented markets ($\text{\texttt{seg}} = 0.95$) is considerably slower, especially in the first three trading sequences (18 years), than for markets with low segmentation ($\text{\texttt{seg}} = 0.05$).

When the true model is \ce{CO2}-driven there is more convergence toward the true model. We believe this is because when models are fit to the historical data, the residuals from the \ce{CO2}-driven model are considerably smaller than those from the TSI-driven model. Thus, projections of future climate change using the TSI model will exhibit significantly larger stochastic noise than the \ce{CO2} models, and this noise makes it more difficult for traders to identify the true model when the true model is TSI.

Traders with higher risk tolerance will price buy and sell orders more aggressively, so they will earn more (or lose less) on completed trades, but will have a greater risk of failing to complete trades. Among traders with the correct model of climate, those with higher risk tolerance  will earn more profit on completed trades (and their counterparties will lose more). A trader's wealth is an important source of information, and we believe this is why greater risk taking enables traders to identify the correct model more quickly. Additionally, the risk of failing to complete trades adds volatility, which also puts information in play. We have observed a similar phenomenon in a very different context, where adding stochastic noise to player decisions in iterated games improves the accuracy with which machine-learning algorithms can identify the players' strategies \shortcite{Nay2015}.

What is not clear to us is why increasing the number of traders slows convergence, even if the number of edges per trader remains fixed. This is a topic for future research.

\section{CONCLUSIONS}

We simulate two alternative climate futures: one where \ce{CO2} is the primary driver of global temperature and one where variations in solar intensity are the primary driver. These represent the two most plausible competing views in the public discourse and our analysis is agnostic about which is ``true''.

Market participation causes traders to converge toward believing the ``true'' climate model under a variety of model parameterizations in a relatively short time: In markets with low segmentation, the number of traders believing the true model rises from 50\% to 75\% in roughly 12 years, and even in highly segmented markets, 75\% convergence is achieved in 18--24 years. Ideally, we would like to compare belief convergence with and without a prediction market but because the only source of climate belief in our current model is market interactions we cannot make this comparison. However, we do use actual temperatures for our model under historical 20th century conditions and observe convergence to the \ce{CO2} model, whereas in the real world there is no convergence of beliefs.

Both the historical and future simulation results suggest that a climate prediction market could be useful for producing broad agreement about the causes of climate change, and could have persuasive power for people who are not persuaded by the overwhelming consensus among climate researchers that greenhouse gases (and especially \ce{CO2}) are responsible for the majority of observed climate change \shortcite{IPCC_WG1_AR5_2013}. We also find that market segmentation has a large effect on the speed of convergence, so transparency and effective communication about the performance of traders with different beliefs will be important.

The fact that rapid convergence to the true model occurs regardless of which model is actually ``true'' may persuade those who doubt the scientific consensus that the market is ideologically neutral, and that the deck is not stacked to produce a pre-determined result.

All code and data for the model is available at \href{https://github.com/JohnNay/predMarket}{github.com/JohnNay/predMarket}. This project is a computational test-bed for public policy design: our code can be extended to test the effects of trading strategies, cognitive models, future climate scenarios, and market designs on the evolution of trader beliefs.

\section*{ACKNOWLEDGEMENTS}

We thank Yevgeniy Vorobeychik for feedback on this project and Manuel Naumovich Velasco for sharing the data from his Total Solar Irradiance prediction.
U.S. National Science Foundation grants EAR-1416964, EAR-1204685, and IIS-1526860 partially supported this research.
All views expressed are the authors' and not the funder's.

\bibliography{prediction_market_wsc}

\begin{thebibliography}{}

\bibitem[\protect\citeauthoryear{Aamodt and Plaza}{Aamodt and
  Plaza}{1994}]{Aamodt1994}
Aamodt, A., and E.~Plaza. 1994.
\newblock ``Case-Based Reasoning: Foundational Issues, Methodological
  Variations, and System Approaches''.
\newblock {\em {AI} Communications\/}~7:39--59.


\bibitem[\protect\citeauthoryear{Archer}{Archer}{2012}]{Archer_2012}
Archer, D. 2012.
\newblock {\em Global Warming: Understanding the Forecast\/}. 2nd ed.
\newblock Wiley.


\bibitem[\protect\citeauthoryear{Beachkofski and Grandhi}{Beachkofski and
  Grandhi}{2002}]{beachkofski_improved_2002}
Beachkofski, B., and R.~Grandhi. 2002.
\newblock ``Improved {D}istributed {H}ypercube {S}ampling''.
\newblock In {\em 43rd {AIAA}/{ASME}/{ASCE}/{AHS}/{ASC} Structures, Structural
  Dynamics, and Materials Conference}. American Institute of Aeronautics and
  Astronautics.

\bibitem[\protect\citeauthoryear{Bloch, Annan, and Bowles}{Bloch
  et~al.}{2010}]{Bloch_2010}
Bloch, D., J.~Annan, and J.~Bowles. 2010.
\newblock ``Cracking the Climate Change Conundrum with Derivatives''.
\newblock {\em Wilmott Journal\/}~2:271--287.


\bibitem[\protect\citeauthoryear{Carnell}{Carnell}{2012}]{carnell_lhs_2012}
Carnell, R. 2012.
\newblock ``{L}atin Hypercube Samples''.
\newblock {R} package, Comprehensive {R} Archive Network.

\bibitem[\protect\citeauthoryear{Carpenter, Gelman, Hoffman, Lee, Goodrich,
  Betancourt, Brubaker, Guo, Li, and Riddell}{Carpenter
  et~al.}{2016}]{Stan2016}
Carpenter, B., A.~Gelman, M.~Hoffman, D.~Lee, B.~Goodrich, M.~Betancourt, M.~A.
  Brubaker, J.~Guo, P.~Li, and A.~Riddell. 2016.
\newblock ``Stan: A Probabilistic Programming Language''.
\newblock {\em Journal of Statistical Software\/}.
\newblock (in press).


\bibitem[\protect\citeauthoryear{Dreber, Pfeiffer, Almenberg, Isaksson, Wilson,
  Chen, Nosek, and Johannesson}{Dreber et~al.}{2015}]{Dreber_2015}
Dreber, A., T.~Pfeiffer, J.~Almenberg, S.~Isaksson, B.~Wilson, Y.~Chen, B.~A.
  Nosek, and M.~Johannesson. 2015.
\newblock ``Using Prediction Markets to Estimate the Reproducibility of
  Scientific Research''.
\newblock {\em Proceedings of the National Academy of
  Sciences\/}~112:15343--15347.


\bibitem[\protect\citeauthoryear{Gelman, Hwang, and Vehtari}{Gelman
  et~al.}{2014}]{Gelman_2014}
Gelman, A., J.~Hwang, and A.~Vehtari. 2014.
\newblock ``Understanding Predictive Information Criteria for Bayesian
  Models''.
\newblock {\em Statistics and Computing\/}~24:997--1016.


\bibitem[\protect\citeauthoryear{{GISTEMP Team}}{{GISTEMP
  Team}}{2016}]{GISTEMP_2016}
{GISTEMP Team} 2016.
\newblock ``{GISS} Surface Temperature Analysis ({GISTEMP})''.
\newblock Technical report, NASA Goddard Institute for Space Studies.

\bibitem[\protect\citeauthoryear{Gode and Sunder}{Gode and
  Sunder}{1993}]{Gode1993}
Gode, D.~K., and S.~Sunder. 1993.
\newblock ``Allocative Efficiency of Markets with Zero-Intelligence Traders:
  Market as a Partial Substitute for Individual Rationality''.
\newblock {\em Journal of Political Economy\/}~101:119--137.


\bibitem[\protect\citeauthoryear{Goel, Reeves, Watts, and Pennock}{Goel
  et~al.}{2010}]{goel_prediction_2010}
Goel, S., D.~M. Reeves, D.~J. Watts, and D.~M. Pennock. 2010.
\newblock ``Prediction Without Markets''.
\newblock In {\em Proceedings of the 11th {ACM} {Conference} on {Electronic}
  {Commerce}}, {EC} '10,  357--366.
\newblock New York, NY, USA: ACM.

\bibitem[\protect\citeauthoryear{Hansen, Ruedy, Sato, and Lo}{Hansen
  et~al.}{2010}]{Hansen_2010}
Hansen, J., R.~Ruedy, M.~Sato, and K.~Lo. 2010.
\newblock ``Global Surface Temperature Change''.
\newblock {\em Reviews of Geophysics\/}~48:RG4004.


\bibitem[\protect\citeauthoryear{Hanson}{Hanson}{2012}]{Hanson2012}
Hanson, R. 2012.
\newblock ``Logarithmic Market Scoring Rules for Modular Combinatorial
  Information Aggregation''.
\newblock {\em The Journal of Prediction Markets\/}:3--15.


\bibitem[\protect\citeauthoryear{Hanson, Oprea, and Porter}{Hanson
  et~al.}{2006}]{Hanson2006}
Hanson, R., R.~Oprea, and D.~Porter. 2006.
\newblock ``Information Aggregation and Manipulation in an Experimental
  Market''.
\newblock {\em Journal of Economic Behavior \& Organization\/}~60:449--459.


\bibitem[\protect\citeauthoryear{Healy, Linardi, Lowery, and Ledyard}{Healy
  et~al.}{2010}]{Healy2010}
Healy, P.~J., S.~Linardi, J.~R. Lowery, and J.~O. Ledyard. 2010.
\newblock ``Prediction Markets: Alternative Mechanisms for Complex Environments
  with Few Traders''.
\newblock {\em Management Science\/}~56:1977--1996.


\bibitem[\protect\citeauthoryear{Horn, Ivens, Ohneberg, and Brem}{Horn
  et~al.}{2014}]{Horn2014c}
Horn, C.~F., B.~S. Ivens, M.~Ohneberg, and A.~Brem. 2014.
\newblock ``Prediction markets---A literature review 2014''.
\newblock {\em The Journal of Prediction Markets\/}~8:89--126.


\bibitem[\protect\citeauthoryear{Hsu}{Hsu}{2011}]{Hsu2011}
Hsu, S.-L. 2011.
\newblock ``A Prediction Market for Climate Outcomes''.
\newblock {\em University of Colorado Law Review\/}~83:179--256.


\bibitem[\protect\citeauthoryear{IPCC}{IPCC}{2013}]{IPCC_WG1_AR5_2013}
IPCC 2013.
\newblock {\em Climate Change 2013: The Physical Science Basis}.
\newblock Cambridge University Press.


\bibitem[\protect\citeauthoryear{Jumadinova and Dasgupta}{Jumadinova and
  Dasgupta}{2011}]{Jumadinova2011a}
Jumadinova, J., and P.~Dasgupta. 2011.
\newblock ``A Multi-Agent System for Analyzing the Effect of Information on
  Prediction Markets''.
\newblock {\em International Journal of Intelligent Systems\/}~26:383--409.


\bibitem[\protect\citeauthoryear{Kahan, Jenkins-Smith, and Braman}{Kahan
  et~al.}{2011}]{Kahan_2011}
Kahan, D.~M., H.~Jenkins-Smith, and D.~Braman. 2011.
\newblock ``Cultural Cognition of Scientific Consensus''.
\newblock {\em Journal of Risk Research\/}~14:147--174.


\bibitem[\protect\citeauthoryear{Klingert and Meyer}{Klingert and
  Meyer}{2012}]{Klingert2012c}
Klingert, F.~M., and M.~Meyer. 2012.
\newblock ``Comparing Prediction Market Mechanisms Using An Experiment-Based
  Multi-Agent Simulation.''.
\newblock In {\em Proceedings 26th European Conference on Modelling and
  Simulation}, edited by\ K.~G. Troitzch, M{\"o}hring, and U.~Lotzmann,
  654--661.

\bibitem[\protect\citeauthoryear{Kolp and Riahi}{Kolp and
  Riahi}{2009}]{rcp_database_2009}
Kolp, P., and K.~Riahi. 2009.
\newblock ``{RCP} Database''.
\newblock Technical report.

\bibitem[\protect\citeauthoryear{Marino, Hogue, Ray, and Kirschner}{Marino
  et~al.}{2008}]{Marino2008}
Marino, S., I.~B. Hogue, C.~J. Ray, and D.~E. Kirschner. 2008.
\newblock ``A Methodology for Performing Global Uncertainty and Sensitivity
  Analysis in Systems Biology''.
\newblock {\em Journal of Theoretical Biology\/}~254:178--196.


\bibitem[\protect\citeauthoryear{Nay and Gilligan}{Nay and
  Gilligan}{2015}]{Nay2015}
Nay, J.~J., and J.~M. Gilligan. 2015.
\newblock ``Data-Driven Dynamic Decision Models''.
\newblock In {\em Proceedings of the 2015 Winter Simulation Conference}, edited
  by\ L.~Yilmaz, W.~Chan, I.~Moon, T.~Roeder, C.~Macal, and M.~Rosetti,
  2752--2763: IEEE Press.

\bibitem[\protect\citeauthoryear{Onta{\~n}{\'o}n and Plaza}{Onta{\~n}{\'o}n and
  Plaza}{2009}]{Ontanon2009}
Onta{\~n}{\'o}n, S., and E.~Plaza. 2009.
\newblock ``Argumentation-Based Information Exchange in Prediction Markets''.
\newblock In {\em Argumentation in Multi-Agent Systems}, edited by\ I.~Rahwan
  and P.~Moraitis, Number 5384 in Lecture Notes in Computer Science,  181--196.
  Springer Berlin Heidelberg.

\bibitem[\protect\citeauthoryear{Pathak, Rothschild, and Dudik}{Pathak
  et~al.}{2015}]{pathak_comparison_2015}
Pathak, D., D.~Rothschild, and M.~Dudik. 2015.
\newblock ``A Comparison of Forecasting Methods: Fundamentals, Polling,
  Prediction Markets, and Experts''.
\newblock {\em Journal of Prediction Markets\/}~9:1--31.


\bibitem[\protect\citeauthoryear{Pujol, Iooss, and Lemaitre}{Pujol
  et~al.}{2014}]{pujol_sensitivity:_2014}
Pujol, G., B.~Iooss, and A.~J. Lemaitre. 2014.
\newblock ``Sensitivity: Sensitivity Analysis''.
\newblock {R} package, Comprehensive {R} Archive Network.

\bibitem[\protect\citeauthoryear{Riahi, Rao, Krey, Cho, Chirkov, Fischer,
  Kindermann, Nakicenovic, and Rafaj}{Riahi et~al.}{2011}]{Riahi_2011}
Riahi, K., S.~Rao, V.~Krey, C.~Cho, V.~Chirkov, G.~Fischer, G.~Kindermann,
  N.~Nakicenovic, and P.~Rafaj. 2011.
\newblock ``RCP 8.5: A Scenario of Comparatively High Greenhouse Gas
  Emissions''.
\newblock {\em Climatic Change\/}~109:33--57.


\bibitem[\protect\citeauthoryear{Saltelli, Chan, and Scott}{Saltelli
  et~al.}{2009}]{saltelli_sensitivity_2009}
Saltelli, A., K.~Chan, and E.~M. Scott. 2009.
\newblock {\em Sensitivity Analysis}.
\newblock Chichester: Wiley.


\bibitem[\protect\citeauthoryear{Set and Selten}{Set and
  Selten}{1998}]{Set1998}
Set, R., and R.~Selten. 1998.
\newblock ``Axiomatic Characterization of the Quadratic Scoring Rule''.
\newblock {\em Experimental Economics\/}~1:43--62.


\bibitem[\protect\citeauthoryear{Soon}{Soon}{2005}]{Soon_2005}
Soon, W. W.-H. 2005.
\newblock ``Variable Solar Irradiance as a Plausible Agent for Multidecadal
  Variations in the Arctic-wide Surface Air Temperature Record of the Past 130
  Years''.
\newblock {\em Geophysical Research Letters\/}~32:L16712.


\bibitem[\protect\citeauthoryear{Tseng, Lin, Lin, Wang, and Li}{Tseng
  et~al.}{2010}]{Tseng2010a}
Tseng, J.-J., C.-H. Lin, C.-T. Lin, S.-C. Wang, and S.-P. Li. 2010.
\newblock ``Statistical Properties of Agent-based Models in Markets with
  Continuous Double Auction Mechanism''.
\newblock {\em Physica A: Statistical Mechanics and its
  Applications\/}~389:1699--1707.


\bibitem[\protect\citeauthoryear{Vandenbergh, Raimi, and Gilligan}{Vandenbergh
  et~al.}{2014}]{Vandenbergh2013f}
Vandenbergh, M.~P., K.~T. Raimi, and J.~M. Gilligan. 2014.
\newblock ``Energy and Climate Change: A Climate Prediction Market''.
\newblock {\em UCLA Law Review\/}~61:1962--2017.


\bibitem[\protect\citeauthoryear{Velasco~Herrera, Mendoza, and
  Velasco~Herrera}{Velasco~Herrera et~al.}{2015}]{Velasco_Herrera_2015}
Velasco~Herrera, V.~M., B.~Mendoza, and G.~Velasco~Herrera. 2015.
\newblock ``Reconstruction and Prediction of the Total Solar Irradiance: From
  the Medieval Warm Period to the 21st century''.
\newblock {\em New Astronomy\/}~34:221--233.


\bibitem[\protect\citeauthoryear{Watanabe}{Watanabe}{2013}]{Watanabe_2013}
Watanabe, S. 2013.
\newblock ``A Widely Applicable Bayesian Information Criterion''.
\newblock {\em Journal of Machine Learning\/}~14:867--897.


\bibitem[\protect\citeauthoryear{Wolfers and Zitzewitz}{Wolfers and
  Zitzewitz}{2006}]{Wolfers2006}
Wolfers, J., and E.~Zitzewitz. 2006.
\newblock ``Prediction Markets in Theory and Practice''.
\newblock Working Paper 12083, NBER.

\end{thebibliography}

\end{document}